\documentclass[11pt]{article}
\usepackage{cite}

\begin{document}

\def \d {{\rm d}}
\def \im {{\rm i}}
\def \eigen {Q}
\def \boldk {\mbox{\boldmath$k$}}
\def \boldl {\mbox{\boldmath$l$}}
\def \boldm {\mbox{\boldmath$m$}}
\def \q {Q}
\def \etta {F} 
\def \A {a}
\def \t {{\Theta}}
\def \k {{\kappa}}
\def \l {{\lambda}}
\def \s {{\sigma}}
\def \tr {{\tilde\rho}}
\def \tv {{\tilde v}}
\def \tz {{\tilde z}}
\def \e {{\epsilon}}
\def \P {{p_\lambda}}
\def \Q {{p_\nu}}

\newcommand{\be}{\begin{equation}}
\newcommand{\ee}{\end{equation}}

\newcommand{\beqn}{\begin{eqnarray}}
\newcommand{\eeqn}{\end{eqnarray}}
\newcommand{\AdS}{anti--de~Sitter }
\newcommand{\AAdS}{\mbox{(anti--)}de~Sitter }
\newcommand{\pa}{\partial}
\newcommand{\pp}{{\it pp\,}-}
\newcommand{\ba}{\begin{array}}
\newcommand{\ea}{\end{array}}

\title{General Kundt spacetimes in higher dimensions}

\author{
 Ji\v{r}\'{\i} Podolsk\'y\thanks{podolsky`AT'mbox.troja.mff.cuni.cz} \ and
 Martin \v{Z}ofka\thanks{zofka`AT'mbox.troja.mff.cuni.cz}
\\ Institute of Theoretical Physics, Faculty of Mathematics and Physics,\\
 Charles University in Prague, V Hole\v{s}ovi\v{c}k\'{a}ch 2, 180 00 Prague 8,  Czech Republic}


\maketitle

\

\abstract{We investigate a general metric of the Kundt class of spacetimes in higher dimensions. Geometrically, it admits a non-twisting, non-shearing and non-expanding geodesic null congruence. We calculate all components of the curvature and Ricci tensors, without assuming any specific matter content, and discuss algebraic types and main geometric constraints imposed by general Einstein's field equations. We explicitly derive Einstein--Maxwell equations, including an arbitrary cosmological constant, in the case of vacuum or possibly an aligned electromagnetic field. Finally, we introduce canonical subclasses of the Kundt family and we identify the most important special cases, namely generalised pp-waves, VSI or CSI spacetimes, and gyratons.

\vspace{.2cm}


\section{Introduction}
\label{intro}

Studies of various aspects of gravity in higher dimensions are now an active research area. In fact, there are already many indications that gravitation in ${D>4}$ exhibits some qualitatively different and even unexpected properties. These can be demonstrated and investigated analytically using exact solutions to Einstein's equations of higher-dimensional general relativity. Such explicit solutions not only illustrate specific physical properties of idealised situations, but may help us to  understand rigorously some of the more general features of the theory.

In standard ${D=4}$ general relativity, there exist many families of spacetimes, as recently summarised in the comprehensive review book \cite{Stephanibook}. Some of them have already been extended to higher dimensions, and great a number of specific exact solutions has been found. However, a more systematic investigation of other interesting families is still desirable.

For example, in our recent works \cite{PodOrt06,OrtPodZof08} we systematically analysed a large class of Robinson--Trautman spacetimes in any dimension, see also \cite{PraPraOrt07}. We found some rather surprising results, in particular that this family is, in a sense, not as rich as in four dimensions. Many types of exact solutions, such as exact gravitational waves of an algebraic type~N, III or~II are completely missing, and even some type~D solutions are absent, e.g., a generalisation of the C-metric that would describe accelerating black holes in higher dimensions.

Geometrically, the Robinson--Trautman class in any dimension $D$ is defined by admitting a geodesic, shear-free, twist-free but expanding null congruence. This invariant definition is based on the optical properties of null geodesic congruences in higher-dimensional spacetimes \cite{FroSto03,PraPraColMil04} (for a review, see \cite{Col08}). A natural counterpart of the Robinson--Trautman family is the Kundt class of spacetimes, which admits a geodesic, shear-free, twist-free and \emph{non-expanding} null congruence.

In fact, the Kundt class is one of the fundamental classes of exact solutions to Einstein's field equations in ${D=4}$ (see chapter~31 in \cite{Stephanibook}). This large group of algebraically special spacetimes contains many particular vacuum solutions. It also admits a cosmological constant, electromagnetic field, pure radiation or other matter fields.

Several important subclasses of Kundt's family in higher dimensions have already been recognised and studied thoroughly. The best-known of these are the pp-waves, see \cite{Col08,Bri25,ColMilPelPraPraZal03,Ort04,Coleyetal04,ColMilPraPra04,ColFusHerPel06} and references therein, which admit a covariantly constant null vector field. There are also VSI and CSI spacetimes \cite{ColMilPelPraPraZal03,Coleyetal04,ColMilPraPra04,ColHerPel06,ColFusHerPel06,Col08}, for which all polynomial scalar invariants constructed from the Riemann tensor and its derivatives vanish and are constant, respectively, and relativistic gyratons
\cite{Bon70,FroFur05,FroIsrZel05,FroZel05,FroZel06,FroLin06,CalLeZor07}, representing the field of a localised spinning source that propagates with the speed of light. Recently, some properties of a more general family of vacuum solutions with a non-twisting multiple WAND, which also includes the Kundt spacetimes, were presented in \cite{PraPra08}. The importance of higher-dimensional Kundt spacetimes in the context of string theory, namely the supersymmetric solutions of supergravity, were summarised in \cite{BraCoHer2008}.

It is the purpose of the present paper to systematically derive the ${D>4}$ Kundt class of solutions and to discuss its main features. Therefore, in section~\ref{sec_geom}, we start with a completely general Kundt metric (\ref{gen_metric}) without assuming any specific matter content. In section~\ref{sec_Ricci} we present all components of the curvature and Ricci tensors for such a general Kundt metric, and in section~\ref{generalfieldeqs} we discuss some properties of the Kundt spacetimes when general Einstein's field equations are applied. In particular, we investigate the constraints imposed on the matter content by the Kundt geometry and we determine its generic algebraic type.

Starting from section~\ref{EinsteinMaxwell}, we confine ourselves to the most important case of vacuum Kundt spacetimes, with a possible cosmological constant $\Lambda$, and we also allow for the presence of an aligned Maxwell field. We derive the explicit form of all Einstein(--Maxwell) equations within such a setting. Our results are summarised in section~\ref{summary} where we also discuss the remaining coordinate freedom and we define canonical subclasses. In section~\ref{previous} we briefly describe the most important subfamilies of these Kundt spacetimes, and we give references to previous works.

\section{Geometry of the Kundt spacetimes}
\label{sec_geom}

The Kundt family of spacetimes in an arbitrary dimension $D$ is  defined geometrically by admitting a geodesic, twist-free, shear-free and non-expanding congruence generated by the null vector field, say $\boldk$. There exist suitable coordinates in which such a metric can be written in the form
\be
 \d s^2=g_{ij}\left(\d x^i+ g^{ri}\d u\right)\left(\d x^j+ g^{rj}\d u\right)-2\,\d u\,\d r-g^{rr}\d u^2 \,.
 \label{geo_metric}
\ee
Indeed, considering a family of null hypersurfaces ${u=const.}$ whose normal (and tangent) is ${\,k_\alpha=- u_{,\alpha}=-\delta^u_\alpha}$ everywhere, the congruence of integral curves of the null vector field $k^\alpha=g^{\alpha\beta}k_\beta$ is then \emph{geodesic and affinely parametrised\,}.
Taking the corresponding affine parameter $r$ along such null congruence as the next coordinate, ${\boldk=\partial_r}$, and introducing ${D-2}$ ``transverse'' spatial coordinates $(x^1, x^2, \ldots , x^{D-2})$ to label the distinct null geodesics of this congruence, we obtain $g^{ur}=-1$ and $g^{uu}=0=g^{ui}$. The remaining specific metric functions in (\ref{geo_metric}), which depend on the coordinates $(x,u,r)$, are to be determined below.\footnote{Here $x$ stands for all the transverse coordinates $x^i$ and  latin indices ${i,j,k,l,m}$ range from 1 to $D-2$.}

Because the generating null vector field $\boldk$ is hypersurface-orthogonal, the congruence is \emph{non-twisting\,}. Writing the spatial part of the metric as
\be
 g_{ij}=p^{-2}\,\gamma_{ij}\qquad \hbox{where} \quad  \det \gamma_{ij}=1\,,
 \label{def of gamma}
\ee
the optical scalars corresponding to shear and expansion, see \cite{FroSto03,PraPraColMil04}, are given by
\beqn
 & & \sigma^2\equiv k_{(\alpha;\beta)}k^{\alpha;\beta}-\frac{1}{D-2}(k^\alpha_{\;;\alpha})^2
   =\frac{1}{4}\gamma^{li}\gamma^{kj}\gamma_{ki,r}\gamma_{lj,r} \,,  \nonumber \label{scalars} \\
 & & \theta\equiv\frac{1}{D-2}k^\alpha_{\;;\alpha}=-(\ln p)_{,r}\,,
\eeqn
respectively. Now, imposing the condition that the congruence is \emph{shear-free} ($\sigma^2=0$) leads to ${\gamma_{ij,r}=0}$ since there always exists a frame in which $\gamma^{ij}$  is diagonal, with strictly positive eingenvalues. Requiring that the congruence is \emph{non-expanding} (${\theta=0}$) leads to ${p_{,r}=0}$. It thus follows that the spatial part of the Kundt metric ${g_{ij}}$ has to be \emph{independent} of the coordinate $\>r$.

The relations between the covariant and contravariant metric coefficients of (\ref{geo_metric}) are\footnote{As usual, $g^{ij}$ denotes the inverse of $g_{ij}$.} ${g^{ri}= g^{ij}g_{uj}}$ and ${g^{rr}=-g_{uu}+g^{ij}g_{ui}g_{uj}}$, or inversely ${ g_{ui}= g_{ij}\,g^{rj}}$, ${g_{uu}=-g^{rr}+g_{ij}\,g^{ri}g^{rj}}$, ${g_{rr}=0=g_{ri} }$. We can thus conclude that any Kundt spacetime in an arbitrary dimension~$D$ can be written in the form
\be
 \d s^2=g_{ij}(x,u)\,\d x^i\d x^j+2\,g_{ui}(x,u,r)\,\d x^i\d u-2\,\d u\,\d r+g_{uu}(x,u,r)\,\d u^2 \,,
 \label{gen_metric}
\ee
see also \cite{ColMilPelPraPraZal03,ColHerPel06,PodOrt06,Col08}.

\section{Curvature tensor for a general Kundt metric}
\label{sec_Ricci}

A straightforward calculation gives the following non-trivial Christoffel symbols for the Kundt metric (\ref{gen_metric}):
\beqn
\Gamma^u_{uu} &=& {\textstyle\frac12} g_{uu,r}\,, \\
\Gamma^u_{ui} &=& {\textstyle\frac12} g_{ui,r}\,, \\
\Gamma^i_{ur} &=& {\textstyle\frac12} g^{ij}g_{uj,r}\,, \\
\Gamma^i_{uu} &=& g^{ij}g_{uj,u}-{\textstyle\frac12} g^{ij}g_{uu,j}-{\textstyle\frac12} g^{ir}g_{uu,r}\,, \\
\Gamma^i_{uj} &=& {\textstyle\frac12} g^{ik} \left( g_{jk,u}+g_{uk,j}-g_{uj,k} \right) -{\textstyle\frac12} g^{ir}g_{uj,r}\,, \\
\Gamma^i_{jk} &=& {\,}^{\rm s}\Gamma^i_{jk} \,,\\
\Gamma^r_{ur} &=& {\textstyle\frac12} g^{ri}g_{ui,r}-{\textstyle\frac12} g_{uu,r}\,, \\
\Gamma^r_{uu} &=& -{\textstyle\frac12}g^{rr}g_{uu,r}-{\textstyle\frac12}g_{uu,u}+g^{ri}g_{ui,u}-{\textstyle\frac12} g^{ri}g_{uu,i}\,, \\
\Gamma^r_{ui} &=& -{\textstyle\frac12}g^{rr}g_{ui,r}-{\textstyle\frac12}g_{uu,i}+{\textstyle\frac12} g^{rj} \left( g_{ij,u}+g_{uj,i}-g_{ui,j} \right), \\
\Gamma^r_{ri} &=& -{\textstyle\frac12}g_{ui,r}\>=-\>\Gamma^u_{ui} \,, \\
\Gamma^r_{ij} &=& {\textstyle\frac12}\left( g_{ij,u}-g_{ui,j}-g_{uj,i} \right)+{\textstyle\frac12} g^{rk} \left( g_{ik,j}+g_{jk,i}-g_{ij,k} \right),
\eeqn
where ${{\,}^{\rm s}\Gamma^i_{jk} }$ is calculated using the spatial metric ${g_{ij}}$ only. Further useful relations are
\beqn
\Gamma^\alpha_{r\alpha} &=& 0\,, \\
\Gamma^\alpha_{u\alpha} &=& \left(\,\ln\sqrt{g}\,\right)_{,u}\,, \\
\Gamma^\alpha_{i\alpha} &=& \Gamma^j_{ij}\>=\>\left(\,\ln\sqrt{g}\,\right)_{,i}\,,
\eeqn
in which we introduced a function
\begin{equation}\label{detg}
g=g(x,u)\equiv \det g_{ij}=-\det g_{\alpha\beta} \,.
\end{equation}
Using these results, a somewhat lengthy calculation then leads to the following components of the Ricci tensor for a general Kundt metric (\ref{gen_metric}):
\beqn
R_{rr} &=& 0\,, \label{Rrr}\\
R_{ri} &=& -{\textstyle\frac12}g_{ui,rr}\,, \label{Rri}\\
R_{ru} &=& -{\textstyle\frac12}g_{uu,rr}+{\textstyle\frac12}(g^{ij}g_{uj,r})_{,i} +{\textstyle\frac14}
       g^{ij}(g_{ui}\,g_{uj})_{,rr}+{\textstyle\frac12} g^{ij}(\,\ln\sqrt{g}\,)_{,i}\,g_{uj,r}\,, \label{Rru}\\
R_{ij} &=& {\,}^{\rm s\!}R_{ij} -{\textstyle\frac12}(\,g_{ui,r}\,g_{uj,r} + g_{ui,rj} + g_{uj,ri}) + {\,}^{\rm s}\Gamma^k_{ij} \,g_{uk,r}\,, \label{Rij}\\
R_{uu} &=& {\textstyle\frac12}(\,g_{uu,rr} - g^{kl}g_{uk,r}\,g_{ul,r})(\,g_{uu} - g^{ij}g_{ui}\,g_{uj})  \nonumber\\
 &&- {\textstyle\frac12}\,g_{uu,r}\left[ (g^{ij}g_{uj})_{,i}+(g^{ij}g_{uj})(\ln\sqrt{g})_{,i} - (\ln\sqrt{g})_{,u} \right]
- {\textstyle\frac12}\!\left(g^{ij}\,g_{ui,r}\,g_{uj}\right)^2  \nonumber\\
 && +{\textstyle\frac12}\,g^{ij}g_{uj,r}\!\left[ \,g_{uu,i}+2g^{kl}g_{uk}(g_{ui,l}- g_{ul,i})\,\right]
+g^{ij}g_{uj}(g_{ui,ru}- g_{uu,ri})   \label{Ruu}\\
 && -{\textstyle\frac12}(g^{ij}g_{uu,j})_{,i} -{\textstyle\frac12}(g^{ij}g_{uu,j}) (\ln\sqrt{g})_{,i} -{\textstyle\frac12}\,g^{ij}g^{kl}g_{uj,l}(g_{uk,i}- g_{ui,k}) \nonumber\\
 &&   +(g^{ij}g_{uj,u})_{,i}+(g^{ij}g_{uj,u})(\ln\sqrt{g})_{,i}
     -{\textstyle\frac14}\,g^{ij}g^{kl}g_{ik,u}\,g_{jl,u} - (\ln\sqrt{g})_{,uu}\,,\nonumber \\
R_{ui} &=& {\textstyle\frac12}\, g^{jk}g_{uk}\,( \,g_{uj,ri}-g_{ui,rj}-g_{ui,r}\,g_{uj,r} )- {\textstyle\frac12}\,g_{uu,ri} + {\textstyle\frac12}\, g^{jk}g_{uk,r}\,g_{uj,i}   \nonumber\\
  &&  + {\textstyle\frac12}\!\left[ \,g^{jk}(\,g_{ij,u}+g_{uj,i}-g_{ui,j}-g_{ui,r}\,g_{uj}) \right]_{,k}
  \nonumber\\
 &&  + {\textstyle\frac12}\!\left[ \,g^{jk}(\,g_{ij,u}+g_{uj,i}-g_{ui,j}-g_{ui,r}\,g_{uj}) \right]\!(\ln\sqrt{g})_{,k}
 \label{Rui}\\
 &&  + {\textstyle\frac12}\, g^{jk}g^{lm}g_{im,k}\left[ (g_{ul,j}-g_{uj,l}) + (g_{ul,r}\,g_{uj}-g_{uj,r}\,g_{ul})\,\right] \nonumber\\
 &&  + {\textstyle\frac12}\,g_{ui,ru}+ {\textstyle\frac12}\,g_{ui,r}(\ln\sqrt{g})_{,u} - (\ln\sqrt{g})_{,ui}- {\textstyle\frac14}\,g^{jk}g^{lm}g_{km,i}\,g_{jl,u} \,,\nonumber
\eeqn
where ${{\,}^{\rm s\!}R_{ij} }$ is calculated using the spatial metric ${g_{ij}}$ only.
The Ricci scalar is thus given by
\beqn
R &=& {\,}^{\rm s\!}R + g_{uu,rr}-(g^{ij}g_{uj,r})_{,i} -{\textstyle\frac12}g^{ij}(g_{ui}\,g_{uj})_{,rr}-g^{ij}(\,\ln\sqrt{g}\,)_{,i}\,g_{uj,r} \nonumber\\
   && - g^{ij}g_{ui}\,g_{uj,rr} -{\textstyle\frac12}g^{ij}(\,g_{ui,r}\,g_{uj,r} + g_{ui,rj} + g_{uj,ri}) + {\,}^{\rm s}\Gamma^k_{ij} \,g^{ij}g_{uk,r}\,. \label{Ricciscalar}
\eeqn
For completeness, we also present all (independent) components of the curvature tensor:
\beqn
R_{rirj} &=& 0\,, \label{Rrirj}\\
R_{rijk} &=& 0\,, \label{Rrijk}\\
R_{riru} &=& -{\textstyle\frac12} g_{ui,rr}\,, \label{Rruri}\\
R_{riuj} &=& {\textstyle\frac12} g_{ui,rj} - {\textstyle\frac12} g_{uk,r} {\,}^{\rm s}\Gamma^k_{ij} + {\textstyle\frac14} g_{ui,r} g_{uj,r}\,, \label{Rriuj}\\
R_{ruru} &=& {\textstyle\frac14}g^{ij}g_{ui,r}g_{uj,r} -{\textstyle\frac12} g_{uu,rr}\,, \label{Rruru}\\
R_{ruij} &=& {\textstyle\frac12} \!\left( \,g_{ui,rj} - g_{uj,ri} \right), \label{Rruij}\\
R_{ruui} &=& {\textstyle\frac12} g_{uu,ri} -{\textstyle\frac12} g_{ui,ru} - {\textstyle\frac14} g_{uj,r}\! \left[\, g^{jk} \left( g_{ik,u} + g_{uk,i} - g_{ui,k} \right) - g^{jr} g_{ui,r}\right], \label{Rruui}\\
R_{ijkl} &=& {\,}^{\rm s\!}R_{ijkl}\,, \label{Rijkl}\\
R_{uijk} &=& {\textstyle\frac12} \!\left( \,g_{uk,ij}+g_{ij,uk}- g_{uj,ik}- g_{ik,uj}\right)
-\Gamma^u_{uk}\Gamma^r_{ij}+\Gamma^u_{uj}\Gamma^r_{ik}  \nonumber\\
 &&  +g_{lm}\!\left(\,\Gamma^l_{uk}\Gamma^m_{ij}-\Gamma^l_{uj}\Gamma^m_{ik}\right) +g_{lu}\!\left(\,\Gamma^u_{uk}\Gamma^l_{ij}-\Gamma^u_{uj}\Gamma^l_{ik}\right)   \,, \label{Ruijk}\\
R_{uiuj} &=& {\textstyle\frac12} \!\left( \,g_{ui,uj}+g_{uj,ui}- g_{uu,ij}- g_{ij,uu}\right)
-\Gamma^r_{ui}\Gamma^u_{uj}-\Gamma^r_{uj}\Gamma^u_{ui}+\Gamma^r_{ij}\Gamma^u_{uu}  \nonumber\\
 &&  +g_{kl}\!\left(\,\Gamma^k_{ui}\Gamma^l_{uj}-\Gamma^k_{ij}\Gamma^l_{uu}\right) +g_{uk}\!\left(\,\Gamma^u_{ui}\Gamma^k_{uj}+\Gamma^u_{uj}\Gamma^k_{ui}-\Gamma^u_{uu}\Gamma^k_{ij}\right) +g_{uu}\Gamma^u_{ui}\Gamma^u_{uj} \,. \label{Ruiuj}
\eeqn

\section{General field equations and algebraic type}
\label{generalfieldeqs}

To determine the specific form of the metric functions in the general line element (\ref{gen_metric}) for the Kundt class of spacetimes, it is now necessary to impose the Einstein field equations ${R_{\alpha\beta}-\frac{1}{2}Rg_{\alpha\beta}+\Lambda g_{\alpha\beta}=8\pi T_{\alpha\beta}}$ with a suitable energy-momentum tensor $T_{\alpha\beta}$. These equations are explicitly rather complicated but, considering the above components of the curvature and metric tensors, some general observations can be made immediately.

First, it follows from equation (\ref{Rrr}) and ${g_{rr}=0}$ that Einstein's field equations for the Kundt class \emph{can only be satisfied provided} ${T_{rr}\equiv T_{\alpha\beta}k^\alpha k^\beta=0}$. This imposes a restriction on the admissible matter content of the spacetime.

If, in addition, ${T_{ri}=0}$ then the field equation corresponding to the component (\ref{Rri}) reduces to a simple relation ${g_{ui,rr}=0}$, which can be directly integrated. In such a case the metric component $g_{ui}$ must be \emph{at most linear} in the coordinate $r$.

From the Einstein equation for (\ref{Rru}), using ${g_{ru}=-1}$ and (\ref{Ricciscalar}), it then follows that $T_{ru}$ must be independent of~$r$. Substituting for $R$ from traced-out Einstein equations ${(D-2)R=2D\Lambda-16\pi \,T_\mu^{\, \mu}}$, we can thus determine the $r$-dependence of the metric function $g_{uu}$. If, for example, the trace ${T_\mu^{\, \mu}}$ of the energy-momentum tensor is independent of $r$, then $g_{uu}$ is at most \emph{quadratic} in $r$.

Subsequently, the remaining field equations corresponding to the components (\ref{Rij})--(\ref{Rui}) have to be used  to fix the undetermined integration functions of $x$ and $u$ (or to rule out a solution due to an inconsistency).

It can also be observed that all higher-dimensional Kundt spacetimes \emph{must necessarily be algebraically special}. They are always at least of \emph{principal type}~I, with ${\boldk=\partial_r}$ being the Weyl aligned null direction (WAND). If ${T_{ri}=0}$ then they are at least of \emph{principal type}~II.

Indeed, using (\ref{Rrr})--(\ref{Ruiuj}), the coordinate components of the Weyl tensor are
\beqn
C_{rirj} &=& 0\,, \label{Crirj}\\
C_{rijk} &=& \frac{1}{2(D-2)} \left(\, g_{ik}\, g_{uj,rr} - g_{ij}\, g_{uk,rr} \right) , \label{Crijk}\\
C_{riru} &=& - \frac{D-3}{2(D-2)}\,g_{ui,rr}  \,, \label{Cruri}
\eeqn
together with much more complicated non-trivial components ${C_{riuj}}$, ${C_{ruru}}$, ${C_{ruij} }$, ${C_{ruui}}$, ${C_{ijkl}}$, ${C_{uijk} }$ and ${C_{uiuj}}$.

Let us introduce a natural null frame
\begin{eqnarray}
\boldm_{(0)}=\boldk  & = & \partial_r\,,\nonumber\\
\boldl  & = & \frac{1}{2} g^{rr}\, \partial_r - \partial_u + g^{ir}\,\partial_i\,,\label{frame}\\
\boldm_{(i)} & = &  p \,\partial_i\,,\nonumber
\end{eqnarray}
for which ${\,g_{\alpha\beta}k^\alpha l^\beta=1\,}$ and ${\,g_{\alpha\beta}\,m_{(i)}^\alpha\, m_{(j)}^\beta=\gamma_{ij}\,}$, with all other scalar products vanishing. (Of course, using a spatial rotation ${\boldm'_{(i)}={X_i}^j\boldm_{(j)}}$ where ${{X_i}^j}$ is a suitable orthogonal matrix the spatial metric $\gamma_{ij}$ can always be diagonalised to $\delta_{ij}$.) Then we obtain the following frame components of the Weyl tensor:
\begin{eqnarray}
C_{(0)(i)(0)(j)} & = & 0\,, \label{frameWeyl1}\\
C_{(0)(i)(j)(k)} & = & \frac{p^3}{2(D-2)} \left(\, g_{ik}\, g_{uj,rr} - g_{ij}\, g_{uk,rr} \right). \label{frameWeyl2}
\end{eqnarray}
According to the classification, reviewed in \cite{Col08}, the Kundt spacetimes are thus necessarily of algebraic type~I, or more special. In the case when ${g_{ij}\, g_{uk,rr} = g_{ik}\, g_{uj,rr}}$, the spacetimes are of type~II (or more special). This occurs, in particular, when ${T_{ri}=0}$ because the corresponding field equations then imply ${g_{ui,rr}=0}$ (see also propositions~1 and~2 of~\cite{OrtPraPra07}).

\section{The Einstein--Maxwell equations}
\label{EinsteinMaxwell}

In the remaining part of this paper we will restrict our attention to spacetimes that are either \emph{vacuum} (${T_{\alpha\beta}=0}$), possibly with a \emph{cosmological constant} $\Lambda$, or those that include a \emph{Maxwell field aligned} with the geometrically privileged null vector $\boldk$ such that
\begin{equation}
F_{\alpha\beta} k^{\beta}=\eigen \, k_\alpha,
\end{equation}
where $\eigen$ is an arbitrary function. The corresponding energy-momentum tensor of the electromagnetic field is
\begin{equation}\label{Energy momentum}
  4\pi\, T_{\alpha\beta} = F_{\alpha \mu} {F_\beta}^\mu- {\textstyle\frac{1}{4}} g_{\alpha\beta} F_{\mu\nu} F^{\mu\nu} .
\end{equation}
In the coordinate system introduced in (\ref{gen_metric}), for ${\boldk=\partial_r}$ there is
\begin{equation}\label{Trivial Implications}
F_{ri}= 0= F^{ui}, \qquad F_{ru}=\eigen=F^{ur},
\end{equation}
with components $F_{ij}$ and $F_{ui}$ (or ${F^{ij}=g^{ik}g^{jl}F_{kl}}$, ${F^{ir}=-\eigen g^{ri}+g^{ij}F_{uj}-g^{rk}g^{ij}F_{kj}}$) still arbitrary. Consequently, ${{F_r}^u={F_r}^i={F_i}^u=0}$ and ${{F_u}^u=-{F_r}^r =\eigen}$, other components are generally non-trivial. In particular, it follows that ${T_{rr} = 0 = T_{ri}\,}$: such spacetimes are of algebraic type~II or more special, see end of section~\ref{generalfieldeqs}. Notice that the trace
${T_\mu^{\, \mu} = -\frac{D-4}{16 \pi} F_{\mu\nu} F^{\mu\nu}}$, which is equal to ${\frac{D-4}{16 \pi} (2\eigen^2 - F_{ij} F^{ij})}$,
is generally non-zero unless ${D=4}$.

The Einstein field equations, re-written as ${R_{\alpha\beta}= \frac{2}{D-2}\,\Lambda\, g_{\alpha\beta} + 8\pi T_{\alpha\beta} - \frac{8\pi}{D-2}\, g_{\alpha\beta}\,T_\mu^{\, \mu}}$, can thus be expressed in the explicit form
\begin{equation}\label{Ricci}
R_{\alpha \beta} = \frac{2}{D-2}\, \Lambda\, g_{\alpha \beta} + 2\, F_{\alpha \mu} {F_\beta}^\mu - \frac{1}{D-2}\, g_{\alpha \beta} F_{\mu\nu} F^{\mu\nu} .
\end{equation}
These will now be calculated for vacuum or electrovacuum, together with the source-free Maxwell equations ${F_{[\alpha\beta;\gamma]}=0}$, ${{F^{\mu\nu}}_{;\nu} = 0}$. The first set of the Maxwell equations is equivalent to ${F_{\alpha\beta,\gamma} + F_{\beta\gamma,\alpha} + F_{\gamma\alpha,\beta}=0}$, while the second one to  ${\sqrt{g}\,{F^{\mu\nu}}_{;\nu} \equiv \big(\sqrt{g}\, F^{\mu\nu}\big)_{,\nu}=0}$.
These yield
\begin{eqnarray}
 F_{ij,r} &=& 0\,, \label{F_ijr}\\
 F_{ui,r} &=& - \eigen_{,i}\,, \label{F_uir}\\
 F_{ij,u} &=& F_{uj,i} - F_{ui,j} \,, \label{F_iju}\\
 F_{[ij,k]}&=& 0 \,, \label{F_ijk}
\end{eqnarray}
and
\begin{eqnarray}
 \eigen_{,r} &=& 0 \,, \label{eigen_r}\\
 \sqrt{g} \;{F^{ir}}_{,r} &=& -\; (\sqrt{g} \; F^{ij})_{,j}\,, \label{eigen_j}\\
 (\sqrt{g} \; F^{ir})_{,i} &=& -(\sqrt{g} \; \eigen)_{,u}\,. \label{eigen_u}
\end{eqnarray}

From (\ref{F_ijr}) we observe that the components $F_{ij}$ are independent of $r$,
\begin{equation}\label{Fij}
F_{ij} = F_{ij} (x,u).
\end{equation}
Using (\ref{eigen_r}), we also find
\begin{equation}\label{Fru}
  F_{ru} = \q(x,u),
\end{equation}
with $\q(x,u)$ arbitrary. Using this result and (\ref{F_uir}), we obtain
\begin{equation}
F_{ui}= -r\, \q_{,i} - \xi_i (x,u), \label{Fui}
\end{equation}
with $\xi_i(x,u)$ arbitrary functions of $x$ and $u$. Substituting now (\ref{Fui}) into  (\ref{F_iju}), we obtain
\begin{eqnarray}
&& F_{ij,u} =  \xi_{i,j} - \xi_{j,i}\> . \label{F_ijuN}
\end{eqnarray}
In particular, taking ${\,F_{ij} \equiv  A_{j,i} - A_{i,j}\,}$ with ${\,A_i=-\int \xi_i \,\d u}$, both (\ref{F_ijk}) and (\ref{F_ijuN}) are satisfied identically.

Thus we found the $r$-dependence of all electromagnetic field components. In particular, the invariant $F_{\mu \nu} F^{\mu\nu}$ of the Maxwell field is \emph{independent} of $r$, and
\begin{equation}
  F_{\mu \nu} F^{\mu\nu} = \etta^2 - 2 \q^2, \label{F_invariant}
\end{equation}
where we defined
\begin{equation}
  \etta^2(x,u) \equiv F_{ik} \,F_{jl}\, g^{ij} g^{kl} .\label{eta2}
\end{equation}
We always have ${\etta^2 \ge 0}$ (with ${\etta^2=0}$ if, and only if, ${F_{ij}=0}$) because, in an orthonormal frame, ${\etta^2=\sum_{\hat i, \hat j}F_{\hat i \hat j}^2\,}$.

Geometry of the Kundt class determines fully the $r$-dependence of the Maxwell field. The non-trivial components of $F_{\mu\nu}$ are explicitly given by (\ref{Fij}), (\ref{Fru}) and (\ref{Fui}), with the remaining constraints (\ref{eigen_j}) and (\ref{eigen_u}).

\subsection{Equations ${R_{rr}=0}$ and ${R_{ri}=0}$}
\label{subsec_Rrr}

It easily follows from the metric structure (\ref{gen_metric}) and expressions (\ref{Trivial Implications}) that two of the Einstein field equations~(\ref{Ricci}) are very simple, namely ${R_{rr}=0}$ and ${R_{ri}=0}$. The former is, in fact, satisfied identically. In view of (\ref{Rri}), the latter reduces to ${\>g_{ui,rr}=0\>}$. This can immediately be integrated yielding
\be
 g_{ui}=e_i(x,u)+ f_i(x,u)\,r ,
  \label{gui}
\ee
where $e_i$ and $f_i$ are arbitrary functions of $x$ and $u$. The $r$-dependence of the metric functions $g_{ui}$ is now determined: interestingly, for any (aligned electro)vacuum Kundt spacetime in an arbitrary dimension these functions are at most linear in the affine parameter $r$. In view of (\ref{frameWeyl2}), these spacetimes are thus algebraically special (at least of type~II). Consequently, the related contravariant metric components are
\be
 g^{ri}=e^i(x,u)+ f^i(x,u)\,r ,
  \label{gri}
\ee
where ${\>e^i\equiv g^{ij}e_j\>}$ and ${\>f^i\equiv g^{ij}f_j\,}$.

\subsection{Equation for ${R_{ur}}$}
\label{subsec_Rur}
The Ricci tensor component $R_{ur}$ for the metric (\ref{gen_metric}) is given by (\ref{Rru}). Using the result obtained above, \
${ 2 F_{u\mu} {F_r}^\mu + \frac{1}{D-2}\,  F_{\mu\nu} F^{\mu\nu}= \frac{1}{D-2}\,\etta^2 +  2\frac{D-3}{D-2}\,\q^2 \,.}$ \
In view of (\ref{gui}), the corresponding field equation (\ref{Ricci}) thus explicitly reads
\begin{equation}\label{Einstein_equation_ur}
  -\frac{1}{2} {g_{uu}}_{,rr} + \varphi = -\frac{2 \Lambda}{D-2} +  \frac{\etta^2+2(D-3)\q^2}{D-2} \,,
\end{equation}
where
\begin{equation}\label{DEF a}
\varphi=\varphi(x,u)\equiv {\textstyle \frac{1}{2}} \left(\,f^i\! f_i +f^i_{\,,i} +  f^i(\ln \sqrt{g}\,)_{,i}\,\right) .
\end{equation}
By a simple integration, the component $g_{uu}$ of the metric is thus determined as
\begin{equation}\label{g_uu}
  g_{uu}= \A(x,u)\, r^2 + b(x,u)\,r + c(x,u)\,,
\end{equation}
where
\begin{equation}\label{DEF AA}
\A(x,u)\equiv  \varphi+\frac{2\Lambda}{D-2} -  \frac{\etta^2+2(D-3)\q^2}{D-2}  \,,
\end{equation}
and $b(x,u)$, $c(x,u)$ are (so far) arbitrary integration functions of $x$ and $u$.

\subsection{Equation for $R_{ij}$}
\label{subsec_Rij}
Since \
${2\, F_{i\mu} {F_j}^\mu - \frac{1}{D-2}\, g_{ij} F_{\mu\nu} F^{\mu\nu}= - \frac{1}{D-2}(\etta^2 - 2 \q^2 )\,g_{ij} +2 F_{ik} F_{jl}\, g^{kl}\,}$ \ and the Ricci tensor component $R_{ij}$ is given by (\ref{Rij}), using   (\ref{gui}) the corresponding field equation (\ref{Ricci}) becomes
\begin{equation}\label{R_ij_tensor}
{\,}^{\rm s\!}R_{ij} = \frac{2\Lambda}{D-2} \, g_{ij} - \frac{\etta^2 - 2 \q^2 }{D-2}\,g_{ij}+2 F_{ik} F_{jl}\, g^{kl}
       + {\textstyle\frac12}(\,f_if_j + f_{i,j} + f_{j,i}) - {\,}^{\rm s}\Gamma^k_{ij} \,f_k \,,
\end{equation}
where ${{\,}^{\rm s\!}R_{ij} }$ and ${{\,}^{\rm s}\Gamma^k_{ij}}$ are the Ricci tensor and the Christoffel symbols of the spatial metric $g_{ij}$, respectively.\footnote{In fact, the last three terms in (\ref{R_ij_tensor}) correspond to (half of) the Lie derivative of the spatial metric $g_{ij}$ with respect to $f^i$, while the last two terms in (\ref{DEF a}) can be re-written as the divergence of $f^i$ with respect to $g_{ij}$, cf. expression~(\ref{covarspatialdiv}) below.} Equations (\ref{R_ij_tensor}) are independent of $r$. Notice that their trace gives
\begin{equation}\label{R_ij_scalar}
  {\,}^{\rm s\!}R = 2 \Lambda + \etta^2 + 2\q^2 + {\textstyle\frac12} f^i\!f_i + f^i_{\,,i} + f^i (\ln \sqrt{g}\,)_{,i}\,,
\end{equation}
where ${{\,}^{\rm s\!}R\>}$ is the spatial Ricci scalar. This relation enables us to re-express the function $\varphi$ as
\begin{equation}\label{ALTER a}
2\,\varphi= {\textstyle \frac{1}{2}}  f^i\! f_i + {\,}^{\rm s\!}R-2\Lambda-(\etta^2 + 2\q^2)\, ,
\end{equation}
and the function $\A$ as
\begin{equation}\label{ALTER aa}
\A= {\textstyle \frac{1}{4}}  f^i\! f_i + {\textstyle \frac{1}{2}}{\,}^{\rm s\!}R-\frac{D-4}{D-2}\,\Lambda-\frac{D(\etta^2 + 2\q^2)+4(D-4)\q^2}{2(D-2)}\, .
\end{equation}
Consequently, the quadratic term in the metric coefficient $g_{uu}$ can be written in the form that does not explicitly contain derivatives of either the functions $f^i$ or the determinant of the metric $g$.

\subsection{Equation for $R_{uu}$}
\label{subsec_Ruu}

Using (\ref{Fru}), (\ref{Fui}) and (\ref{F_invariant}), the right-hand side of the  ${uu}$ component of Einstein's equations (\ref{Ricci}) becomes
\begin{eqnarray}\label{T_uu}
&&\frac{2\Lambda}{D-2}\,g_{uu} + 2\, F_{u\mu} {F_u}^\mu - \frac{1}{D-2}\, g_{uu} F_{\mu\nu} F^{\mu\nu}  =
\left[ \,\A(\A -\varphi) + 2\,g^{ij}(\q_{,i} + \q f_i) (\q_{,j} + \q f_j)\,   \right] r^2 \nonumber\\
&&\qquad+ \left[ \,\A (b-\varphi) + 4\,(\q_{,i} + \q f_i) (\xi^i + \q\, e^i)\,   \right] r
  + \left[ \,\A (c-\varphi) + 2\,(\xi_i + \q\, e_i) (\xi^i + \q\, e^i)\,   \right]  .
\end{eqnarray}
The general form of the Ricci tensor component $R_{uu}$ is given by (\ref{Ruu}). By substituting from (\ref{gui}), (\ref{g_uu}) and comparing the coefficients of different powers of $r$, we obtain the following three equations, namely
\begin{eqnarray}
&&{\textstyle
\frac{1}{2}(g^{ij}\A_{,j})_{||i}  + \frac{3}{2}f^i\A_{,i} + \A (\varphi+\frac{1}{2}f^i\! f_i) }\label{Einstein_equation_uu_order_two}\\
&&\quad {\textstyle
-\frac{1}{2}g^{ij}g^{kl}f_{j,l}(f_{i,k}-f_{k,i})}=
{\textstyle
-2\,g^{ij}(\q_{,i} + \q f_i) (\q_{,j} + \q f_j)}
\nonumber
\end{eqnarray}
for $r^2$,
\begin{eqnarray}
&&{\textstyle
- \frac{1}{2}(g^{ij}b_{,j})_{||i}  + \frac{1}{2}f^i b_{,i}  -2\,e^i\A_{,i}-\A \big(\,e^i_{\,||i}  + 2\,e^if_i- (\ln \sqrt{g}\,)_{,u}  \big)
 }\label{Einstein_equation_uu_order_one}\\
&&\quad {\textstyle
+f^ie^j(f_{i,j}-f_{j,i})+g^{ij}g^{kl}e_{j,l}(f_{i,k}-f_{k,i})}\nonumber\\
&&\quad {\textstyle
+(g^{ij}f_{j,u})_{||i} +f^i(f_{i,u}-b_{,i})}
=
{\textstyle
4(\q_{,i} + \q f_i) (\xi^i + \q\, e^i)}
\nonumber
\end{eqnarray}
for $r^1$, and
\begin{eqnarray}
&&{\textstyle
- \frac{1}{2}(g^{ij}c_{,j})_{||i} + \frac{1}{2}f^i c_{,i} +\frac{1}{2} f^i_{\,||i}\,c -\A \,e^ie_{i}-\frac{1}{2}b\big(\,e^i_{\,||i} - (\ln \sqrt{g}\,)_{,u}\big)
 }\label{Einstein_equation_uu_order_zero}\\
&&\quad {\textstyle
+\frac{1}{2}(f^i\!f_i)(e^je_j)-\frac{1}{2}(f^ie_i)^2+f^ie^j(e_{i,j}-e_{j,i})+\frac{1}{2}g^{ij}g^{kl}e_{j,l}(e_{i,k}-e_{k,i})}\nonumber\\
&&\quad {\textstyle
+(g^{ij}e_{j,u})_{||i}   +e^i(f_{i,u}-b_{,i})  -\frac{1}{4}g^{ij}g^{kl}g_{ik,u}\,g_{jl,u}- (\ln \sqrt{g}\,)_{,uu}} =
{\textstyle
 2(\xi_i + \q\, e_i) (\xi^i + \q\, e^i)}\nonumber
\end{eqnarray}
for $r^0$, respectively. Here, we used the abbreviation
\begin{equation}
\psi^i_{\,||i}\equiv {\psi^i}_{,i}+\psi^i\,(\ln\sqrt{g}\,)_{,i}
\label{covarspatialdiv}
\end{equation}
for the covariant spatial divergence of any quantity $\psi^i$.

\subsection{Equation for $R_{ui}$}

\label{subsec_Rui}

Finally, the general form of the Ricci tensor component $R_{ui}$ for the metric (\ref{gen_metric}) is (\ref{Rui}). Substituting from (\ref{gui}), (\ref{g_uu}) and comparing the coefficients of different powers of $r$ with the corresponding terms on the right-hand side of the Einstein equations (\ref{Ricci}), the following two relations are obtained:
\begin{eqnarray}
&&{\textstyle \frac{1}{2}}f^j(2f_{j,i}-f_{i,j}-f_i f_j)  - \varphi_{,i} -\frac{(\etta^2)_{,i}+2(D-3)(\q^2)_{,i}}{D-2}
 \nonumber\\
&&\quad
 +{\textstyle \frac{1}{2}}\!\left[g^{jk}(f_{j,i}-f_{i,j} -f_i f_j)\right]_{,k}
 +{\textstyle \frac{1}{2}}\left[g^{jk}(f_{j,i}-f_{i,j} -f_i f_j)\right]\,(\ln \sqrt{g}\,)_{,k} \label{Einstein_equation_ui_order_oneFULL}\\
&&\quad +{\textstyle \frac{1}{2}}g^{jk}g^{lm}g_{im,k}(f_{l,j}-f_{j,l})\ = \
\left(\frac{2\Lambda}{D-2} -  \frac{\etta^2-2\q^2}{D-2} \right)\!f_i + 2\q \q_{,i} - 2 F_{ij}(\q f^j + g^{jk}\q_{,k} )\nonumber
\end{eqnarray}
for $r^1$, and
\begin{eqnarray}
&&{\textstyle \frac{1}{2}}\left[g^{jk}(g_{ij,u}+e_{j,i}-e_{i,j} -f_i e_j)\right]_{,k}
 +{\textstyle \frac{1}{2}}\left[g^{jk}(g_{ij,u}+e_{j,i}-e_{i,j} -f_i e_j)\right]\,(\ln \sqrt{g}\,)_{,k}\nonumber\\
&&\quad
 +{\textstyle \frac{1}{2}}e^j(f_{j,i}-f_{i,j}-f_i f_j)  - {\textstyle \frac{1}{2}}b_{,i} + {\textstyle \frac{1}{2}} f^je_{j,i} +{\textstyle \frac{1}{2}}g^{jk}g^{lm}g_{im,k}(e_{l,j}-e_{j,l}+f_l\,e_j-f_j\,e_l) \nonumber\\
&&\quad {\textstyle
+\frac{1}{2}f_{i,u}+\frac{1}{2}f_i(\ln \sqrt{g}\,)_{,u} - (\ln \sqrt{g}\,)_{,ui}-\frac{1}{4}g^{jk}g^{lm}g_{km,i}\,g_{jl,u}} \label{Einstein_equation_ui_order_zeroFULL}\\
&&\hspace{46mm} \ = \
\left(\frac{2\Lambda}{D-2} -  \frac{\etta^2-2\q^2}{D-2} \right)\!e_i + 2\q\, \xi_i - 2 F_{ij}(\q \,e^j + g^{jk} \xi_k)\quad
\nonumber
\end{eqnarray}
for $r^0$.

\section{Summary of the results}
\label{summary}

By applying the Einstein--Maxwell field equations, we have obtained above a complete family of Kundt's spacetimes in an arbitrary dimension $D$ which are either vacuum or contain an aligned electromagnetic field. A non-vanishing cosmological constant $\Lambda$ is also allowed. All such metrics can be written in the form
\be
 \d s^2=g_{ij}\,\d x^i\d x^j+2\,(e_i+ f_i\,r )\,\d x^i\d u-2\,\d u\,\d r+(\A\,r^2 + b\,r + c)\,\d u^2 ,\,
 \label{gen_metric gen}
\ee
where $g_{ij}$, $e_i$, $f_i$, $\A$, $b$ and $c$ are functions of $x$ and $u$ only --- see equations (\ref{gen_metric}), (\ref{gui}), (\ref{g_uu}) and (\ref{ALTER aa}) or, equivalently, by (\ref{DEF AA}), with the function $\varphi$ given by (\ref{DEF a}) or (\ref{ALTER a}). The spatial coordinates ${x\equiv(x^1, x^2, \ldots , x^{D-2})}$ span the transverse space, $u$~labels the family of null surfaces, and $r$~is the affine parameter along the geodesic, twist-free, shear-free and non-expanding congruence generated by the null vector field ${\boldk=\partial_r}$, which is normal to ${u=const}$. Such a vector represents a multiple WAND, and all these spacetimes are of type~II, or more special.

The functions $g_{ij}$, $e_i$, $f_i$, $\A$, $b$ and $c$ in the metric (\ref{gen_metric gen}) are constrained by the remaining Einstein equations, namely (\ref{R_ij_tensor}), (\ref{Einstein_equation_uu_order_two}), (\ref{Einstein_equation_uu_order_one}), (\ref{Einstein_equation_uu_order_zero}), (\ref{Einstein_equation_ui_order_oneFULL}) and (\ref{Einstein_equation_ui_order_zeroFULL}). These equations also contain the non-trivial Maxwell field variables $F_{\mu\nu}$, namely $F_{ij}$, ${\q=F_{ru}}$ and $\xi_i$, such that ${F_{iu}= r\, \q_{,i} + \xi_i (x,u)}$, which all depend only on $x$ and $u$. The corresponding electromagnetic field thus can be written as
\begin{eqnarray}
  \mbox{\boldmath$F$} = \q\, \d r \wedge \d u + (r\, \q_{,i} + \xi_i )\, \d x^i \wedge \d u +{\textstyle\frac{1}{2}}F_{ij}\, \d x^i \wedge \d x^j \>, \label{Maxwell2form}
\end{eqnarray}
Recall also that ${\etta^2(x,u) \equiv F_{ik} \,F_{jl}\, g^{ij} g^{kl} }$. The remaining Maxwell equations are (\ref{eigen_j}) and (\ref{eigen_u}) which, if expanded in the powers of $r$ using the previous results, in particular (\ref{gri}), yield the following two equations in ${D>4}$:
\begin{eqnarray}
(\sqrt{g}\,\q)_{,u}  &=& \left[\,\sqrt{g}\,\left(\q \,e^i - g^{ij} F_{jk}\,e^k + g^{ij} \xi_j\right)\, \right]_{,i}\,, \label{c6}\\
\left(\sqrt{g}\,g^{ik}g^{jl}F_{kl}\right)_{,j}&=& \>\sqrt{g}\,\left(\q f^i - g^{ij} F_{jk}f^k + g^{ij} Q_{,j}\right) \,. \label{c1}
\end{eqnarray}
The relations (\ref{c6}), (\ref{c1}), together with the constraints (\ref{F_ijk}), (\ref{F_ijuN}), which can be solved --- for example --- by taking ${\,F_{ij}=A_{j,i} - A_{i,j}\,}$ where ${\,A_i=-\int \xi_i \,\d u}$, place restrictions on the admissible electromagnetic fields (\ref{Fij})--(\ref{Fui}) in the Kundt family of spacetimes.

Note that such electromagnetic fields can be null when ${\etta^2 = 2 \q^2}$, see the invariant (\ref{F_invariant}). This property is different from the case of higher-dimensional shear-free expanding spacetimes, in particular of the Robinson--Trautman family, which do \emph{not} admit aligned null Maxwell fields \cite{OrtPodZof08}.

\subsection{Coordinate and gauge freedom}
\label{coordfreedom}

The general Kundt metric (\ref{gen_metric}) is left invariant under the coordinate transformations
\be
 x^i=x^i(\tilde x,\tilde u) , \qquad u=u(\tilde u), \qquad r=\frac{\tilde r}{\dot u(\tilde u)} + \rho(\tilde x,\tilde u) \,,
 \label{freedom}
\ee
with $\dot u$ denoting the derivative of $u(\tilde u)$ (cf. \cite{Stephanibook}, section~2 in \cite{ColFusHerPel06}, or section 4 in \cite{ColHerPel06}). This clearly does not change the foliation to a family of null hypersuperfaces ${u=const.}$ nor the affine character of the parameter~$r$. In particular, the form of the metric (\ref{gen_metric gen}) is invariant under (\ref{freedom}) with the metric functions changing as
\begin{eqnarray}
\tilde{g}_{kl}&=& g_{ij}\,\frac{\partial x^i}{\partial \tilde x^k}\frac{\partial x^j}{\partial \tilde x^l}\> , \label{free1}\\
\tilde{e}_k   &=& (e_i+f_i\,\rho)\,\dot u\frac{\partial x^i}{\partial \tilde x^k}-\dot u\frac{\partial\rho}{\partial \tilde x^k}+g_{ij}\,\frac{\partial x^i}{\partial \tilde x^k}\frac{\partial x^j}{\partial \tilde u}\>, \label{free2}\\
\tilde{f}_k   &=& f_i\frac{\partial x^i}{\partial \tilde x^k}\>, \label{free3}\\
\tilde{a}  &=& a\>, \label{free4}\\
\tilde{b}  &=& (b+2a\rho)\,\dot u+2 f_i\frac{\partial x^i}{\partial \tilde u}+2\frac{\ddot u}{\dot u}\>, \label{free5}\\
\tilde{c}  &=& (c+b\rho+a\rho^2)\,\dot u^2-2\dot u\frac{\partial\rho}{\partial \tilde u}+2(e_i+f_i\,\rho)\,\dot u\frac{\partial x^i}{\partial \tilde u}+g_{ij}\,\frac{\partial x^i}{\partial \tilde u}\frac{\partial x^j}{\partial \tilde u}\>. \label{free6}
\end{eqnarray}
The transformation (\ref{freedom}) also induces a change in the electromagnetic field (\ref{Maxwell2form}), namely
\begin{eqnarray}
\tilde{\q}  &=& \q\> , \label{Maxfree1}\\
\tilde{\q}_{,\tilde k}   &=& \q_{,i}\,\frac{\partial x^i}{\partial \tilde x^k}\>, \label{Maxfree2}\\
\tilde{\xi}_k   &=& \left(\xi_i\,\frac{\partial x^i}{\partial \tilde x^k}+\q\,\frac{\partial\rho}{\partial \tilde x^k}+\rho\,\q_{,i}\,\frac{\partial x^i}{\partial \tilde x^k}\right)\dot u + F_{ij}\,\frac{\partial x^i}{\partial \tilde x^k}\frac{\partial x^j}{\partial \tilde u}\>, \label{Maxfree3}\\
\tilde{F}_{kl}&=& F_{ij}\,\frac{\partial x^i}{\partial \tilde x^k}\frac{\partial x^j}{\partial \tilde x^l}\>, \label{Maxfree4}
\end{eqnarray}
which implies ${\tilde{\etta}^2=\etta^2}$. Using this coordinate freedom, a simplification of the metric and/or of the Maxwell field can be achieved. For example, it is (generally) possible to remove the functions~$b$ and~$c$, or to simplify the functions $e_i$ and $f_i$. In particular:

\begin{itemize}

\item Provided ${a\not=0}$, the function $b$ can always be removed  by choosing ${\,\rho=-b/2a}$ in  (\ref{freedom}), keeping $x$ and $u$ unchanged, see (\ref{free5}).

\item Alternatively, in view of (\ref{free6}), it is possible to remove the function $c$ if $\rho$ is taken to be a solution of the differential equation${\>2\,\rho_{,\tilde u}=a\rho^2+b\rho+c}$.

\item We may set, at least \emph{locally}, ${e_i=0}$ by the transformation
${x^i(\tilde x, u) = -\int e^i(\tilde x,u)\,\d u}$.
This is regular only when the determinant $\det J^i_{\,k}$ of the Jacobi matrix
${J^i_{\,k}\equiv\frac{\partial x^i}{\partial\tilde x^k}}$ is non-vanishing. In the degenerate case $\det J^i_{\,k}=0$, we may alternatively remove $e^i$, for example, by applying the transformation
${x^i= -\int e^i(\tilde x,u)\,\d u - \lambda\,\tilde x^i}$.
This is now clearly regular provided ${\lambda\not=0}$ is any real parameter different from the eigenvalues of the matrix $J^i_{\,k}$ in the neighbourhood of a given point.

\item Considering (\ref{free3}), the functions ${f_i}$ can also be simplified using the coordinate freedom (\ref{freedom}). For example, at any given point, it is always possible to apply a suitable ``rotation'' to achieve (say) ${f_1\not=0}$ and ${f_i=0}$ for ${i=2, 3, \dots, D-2}$. However, to obtain such a simplification in a local neighbourhood or even globally, additional conditions must be satisfied.

\item It is sometimes possible to remove the function $\xi_i$ in the electromagnetic component $F_{iu}$ by transforming $x$ so that ${F_{ij}\frac{\partial x^j}{\partial \tilde u}=-\xi_i}$, with $u=\tilde u$ and ${\rho=0}$, see (\ref{Maxfree3}).

\end{itemize}

Of course, these transformations are not mutually independent. Moreover, the particular use of the coordinate freedom is very different for specific subclasses of the Kundt spacetimes.

\subsection{Canonical subclasses}
\label{subclasses}

It also follows immediately from the coordinate freedom (\ref{freedom}) discussed above that
\begin{equation}
f^i\! f_i \equiv g^{ij}f_i f_j  \quad\hbox{ is an \emph{invariant}\,,}
\label{invariant}
\end{equation}
i.e., ${\tilde{g}^{kl}\tilde{f}_k \tilde{f}_l=g^{ij}f_i f_j }$, see the transformation properties (\ref{free1}) and (\ref{free3}). Moreover, due to the positivity of the spatial metric $g_{ij}$, the expression $f^i\! f_i$ can not be negative. This fact can conveniently be used for a natural canonical classification of a general family of the Kundt spacetimes in any dimension.

Namely, it is possible to distinguish \emph{two separate cases}: ${\,f^i\! f_i=0\,}$ and ${\,f^i\! f_i>0\,}$. In the first case, it follows that ${f_i=0}$ for all ${i=1,2,\ldots, D-2}$, see section~\ref{subclasse1} below. In the second case it is sometimes possible to employ the coordinate freedom (\ref{free3}) to obtain a specific simpler form of the functions ${f_i}$, see section~\ref{subclasse2}.

Note also that the functions $f_i$ are directly related to the coefficients
\be
 \tau_i \equiv k_{\alpha;\beta}\,m_{(i)}^\alpha \,l^\beta
 \label{deftaui}
\ee
(which are higher-dimensional analogues of the real and imaginary parts of the Newmann--Penrose spin coefficient $\tau$) with respect to the frame (\ref{frame}) that is naturally adapted to the coordinates of the Kundt metric (\ref{gen_metric gen}). In \cite{PraPraColMil04,Col08,ColFusHerPel06} such coefficients are denoted ${L_{i1}=L_{1i}}$. Using the fact that ${k_{\alpha;\beta}=\frac12g_{\alpha\beta,r}}$, it can easily be shown that
\be
 \tau_i=-\textstyle{\frac12}p\,f_i\,.
 \label{taui}
\ee
In the Kundt class of spacetimes these quantities are invariant under null rotations with respect to a fixed WAND $\boldk$, see \cite{OrtPraPra07} with ${L_{i0}=0=L_{ij}}$, while under spatial rotations ${\boldm'_{(i)}=X_i^j\boldm_{(j)}}$ (where ${X_i^j}$ is an orthogonal matrix) they transform simply as ${\tau'_i=X_i^j\tau_j}$.

Note finally that the above invariant (\ref{invariant}), which occurs (for example) in the explicit expression for the metric function $a$ in (\ref{ALTER aa}), reads
\be{\textstyle \frac{1}{4}}  f^i\! f_i =p^{-2}g^{ij}\,\tau_i \tau_j =\gamma^{ij}\,\tau_i \tau_j\,.
 \label{taubartau}
\ee
In ${D=4}$ this becomes simply ${\,\tau\bar\tau}$.

\subsubsection{The case ${\,f^i\! f_i=0\,}$}
\label{subclasse1}
Clearly, the simplest subclass of the Kundt family of spacetimes is that for which \emph{all the functions $f_i$ vanish\,}, ${f_i=0}$ for all $i$. This occurs if, and only if, ${\,f^i\! f_i=0\,}$ and the corresponding metric and the field equations simplify considerably. Indeed, metric (\ref{gen_metric gen}) reduces to
\be
 \d s^2=g_{ij}\,\d x^i\d x^j+2\,e_i\,\d x^i\d u-2\,\d u\,\d r+(\A\,r^2 + b\,r + c)\,\d u^2 ,\,
 \label{gen_metric fnula}
\ee
where (since ${\varphi=0}$)
\begin{equation}\label{sDEF AA}
\A = \frac{2\Lambda}{D-2} -  \frac{\etta^2+2(D-3)\q^2}{D-2}  \,,
\end{equation}
and the equations (\ref{R_ij_tensor})--(\ref{Einstein_equation_ui_order_zeroFULL}) become
\begin{eqnarray}
&&{\,}^{\rm s\!}R_{ij} = \frac{2\Lambda}{D-2} \, g_{ij} - \frac{\etta^2 - 2 \q^2 }{D-2}\,g_{ij}+2 F_{ik} F_{jl}\, g^{kl}\,,\label{sR_ij_tensor}\\
&&{\textstyle \frac{1}{2}(g^{ij}\A_{,j})_{||i}} = {\textstyle -2\,g^{ij}\q_{,i}\q_{,j}}\,,
\label{sEinstein_equation_uu_order_two}\\
&&{\textstyle \frac{1}{2}(g^{ij}b_{,j})_{||i} + 2\,e^i\A_{,i} + \A \big(\,e^i_{\,||i} - (\ln \sqrt{g}\,)_{,u}  \big) }=
-{\textstyle 4\q_{,i}(\xi^i + \q\, e^i)}\,,\label{sEinstein_equation_uu_order_one}\\
&&{\textstyle \frac{1}{2}}(g^{ij}c_{,j})_{||i} +\A\,e^ie_{i}+{\textstyle\frac{1}{2}}b\big(\,e^i_{\,||i} - (\ln \sqrt{g}\,)_{,u}\big)
 -{\textstyle\frac{1}{2}}g^{ij}g^{kl}e_{j,l}(e_{i,k}-e_{k,i})\nonumber\\
&&\qquad -{\textstyle(g^{ij}e_{j,u})_{||i} + e^ib_{,i} +\frac{1}{4}g^{ij}g^{kl}g_{ik,u}\,g_{jl,u}+ (\ln \sqrt{g}\,)_{,uu}} =
-{\textstyle  2(\xi_i + \q\, e_i) (\xi^i + \q\, e^i)}\,,\hskip10mm\label{sEinstein_equation_uu_order_zero}\\
&&\frac{(\etta^2)_{,i}+2(D-3)(\q^2)_{,i}}{D-2} =
 -2\q \q_{,i} + 2 F_{ij}g^{jk}\q_{,k} \,, \label{sEinstein_equation_ui_order_oneFULL}\\
&&{\textstyle \frac{1}{2}}\left[g^{jk}(g_{ij,u}+e_{j,i}-e_{i,j})\right]_{,k}
 +{\textstyle \frac{1}{2}}\left[g^{jk}(g_{ij,u}+e_{j,i}-e_{i,j})\right]\,(\ln \sqrt{g}\,)_{,k}\nonumber\\
&&\qquad
 - {\textstyle \frac{1}{2}}b_{,i}  +{\textstyle \frac{1}{2}}g^{jk}g^{lm}g_{im,k}(e_{l,j}-e_{j,l})  {\textstyle
 - (\ln \sqrt{g}\,)_{,ui}-\frac{1}{4}g^{jk}g^{lm}g_{km,i}\,g_{jl,u}} \label{sEinstein_equation_ui_order_zeroFULL}\\
&&\qquad =
\left(\frac{2\Lambda}{D-2} -  \frac{\etta^2-2\q^2}{D-2} \right)\!e_i + 2\q\, \xi_i - 2 F_{ij}(\q \,e^j + g^{jk} \xi_k)\,.
\nonumber
\end{eqnarray}
These equations are quite complex, but the coordinate freedom (\ref{freedom}), implying (\ref{free1})--(\ref{Maxfree4}), can be used for some further simplification. For example, when ${a\not=0}$, the function $b$ can always be removed by choosing ${\,\rho=-b/2a}$. In view of (\ref{sDEF AA}), this is always possible if ${\Lambda<0}$.

Notice also that the equation (\ref{sEinstein_equation_uu_order_two})  for $\A$ can be re-written in the invariant form
\be
 \bigtriangleup \A = -4\,|\nabla \q|^2 \leq 0\,, \label{Laplacian on A}
\ee
where ${\bigtriangleup \A \equiv \A^{||i}_{\,\,||i}}$ is the Laplacian on the Riemannian transverse space. Due to (\ref{Laplacian on A}) and the fact that $g_{ij}$ is positive-definite, the function $-\A$ is subharmonic. 

Specifically, by Bochner's ``maximum principle'', on a \emph{compact} Riemannian manifold, a subharmonic function is constant \cite{KobNom2,Bochner48}. Therefore, $\A$ is independent of $x^i$, and so is then~$\q$. From (\ref{sEinstein_equation_ui_order_oneFULL}) it then follows that $F^2$ has the same property, too. Taking the trace of (\ref{sR_ij_tensor}), we conclude that ${\,}^{\rm s\!}R$ is also independent of the transverse spatial coordinates $x$ and the subspace is of constant Ricci curvature, which can only depend on the variable $u$. Maxwell equation (\ref{c1}) reduces to an effective Maxwell equation on the transverse space.

In the \emph{vacuum} case it immediately follows from (\ref{sR_ij_tensor}) and (\ref{sDEF AA}) that
\begin{equation}
{\,}^{\rm s\!}R_{ij} = \A \, g_{ij}\,, \qquad \hbox{where} \quad \A = \frac{2\Lambda}{D-2}  \,, \label{sRijvacuum}
\end{equation}
so that the Riemannian transverse space is an Einstein space.

\subsubsection{The case ${\,f^i\! f_i>0\,}$}
\label{subclasse2}

When the invariant (\ref{invariant}) does not vanish, some of the functions $f_i$ are non-zero. The Einstein--Maxwell equations (\ref{R_ij_tensor})--(\ref{Einstein_equation_ui_order_zeroFULL}) to be solved are then, in general, much more involved. However, it can be observed that a simplification occurs if ${f_{i,j}-f_{j,i}=0}$ (i.e., ${f_{\,[i\,||\,j]}=0}$), so that $f_i$ can locally be written as a gradient. Interestingly, this special case is physically important, as it contains all the VSI spacetimes (see equation (22) in \cite{ColFusHerPel06}). Let us now describe such a particular subcase explicitly.

\subsubsection*{The subcase ${f_{i,j}=f_{j,i}}$}

If ${f_{i,j}=f_{j,i}}$ for all spatial indices $i,j$, the general field equations (\ref{R_ij_tensor})--(\ref{Einstein_equation_ui_order_zeroFULL}) reduce to
\begin{eqnarray}
&&{\,}^{\rm s\!}R_{ij} = \frac{2\Lambda}{D-2} \, g_{ij} - \frac{\etta^2 - 2 \q^2 }{D-2}\,g_{ij}+2 F_{ik} F_{jl}\, g^{kl}
       + {\textstyle\frac12}f_if_j + f_{i,j} - {\,}^{\rm s}\Gamma^k_{ij} \,f_k \,,\label{gR_ij_tensor}\\
&&{\textstyle \frac{1}{2}(g^{ij}\A_{,j})_{||i}  + \frac{3}{2}f^i\A_{,i} + \A (\varphi+\frac{1}{2}f^i\! f_i) }= {\textstyle -2\,g^{ij}(\q_{,i} + \q f_i) (\q_{,j} + \q f_j)}\,,\label{gEinstein_equation_uu_order_two}\\
&&{\textstyle
\frac{1}{2}(g^{ij}b_{,j})_{||i}  - \frac{1}{2}f^i b_{,i}  +2\,e^i\A_{,i}+\A \big(\,e^i_{\,||i}  + 2\,e^if_i- (\ln \sqrt{g}\,)_{,u}  \big)
 }\nonumber\\
&&\qquad {\textstyle
-(g^{ij}f_{j,u})_{||i} -f^i(f_{i,u}-b_{,i})}=-{\textstyle 4(\q_{,i} + \q f_i) (\xi^i + \q\, e^i)}\,,\label{gEinstein_equation_uu_order_one}\\
&&{\textstyle
\frac{1}{2}(g^{ij}c_{,j})_{||i} - \frac{1}{2}f^i c_{,i} -\frac{1}{2} f^i_{\,||i}\,c +\A \,e^ie_{i}+\frac{1}{2}b\big(\,e^i_{\,||i} - (\ln \sqrt{g}\,)_{,u}\big)  }\nonumber\\
&&\qquad {\textstyle
-\frac{1}{2}(f^i\!f_i)(e^je_j)+\frac{1}{2}(f^ie_i)^2-f^ie^j(e_{i,j}-e_{j,i})-\frac{1}{2}g^{ij}g^{kl}e_{j,l}(e_{i,k}-e_{k,i})}\label{gEinstein_equation_uu_order_zero}\\
&&\qquad {\textstyle
-(g^{ij}e_{j,u})_{||i}   -e^i(f_{i,u}-b_{,i})  +\frac{1}{4}g^{ij}g^{kl}g_{ik,u}\,g_{jl,u} +(\ln \sqrt{g}\,)_{,uu}} =
{\textstyle -2(\xi_i + \q\, e_i) (\xi^i + \q\, e^i)}\,,\nonumber\\
&&{\textstyle \varphi_{,i}-\frac{1}{2}}f^j(f_{j,i}-f_i f_j)  +{\textstyle \frac{1}{2}}\!\left(f_i f^j\right)_{,j}
 +{\textstyle \frac{1}{2}}\left(f_i f^j\right)\,(\ln \sqrt{g}\,)_{,j} +\frac{(\etta^2)_{,i}+2(D-3)(\q^2)_{,i}}{D-2} \nonumber\\
&&\qquad   = -\left(\frac{2\Lambda}{D-2} -  \frac{\etta^2-2\q^2}{D-2} \right)\!f_i - 2\q \q_{,i} + 2 F_{ij}(\q f^j
+ g^{jk}\q_{,k} )\,,\label{gEinstein_equation_ui_order_oneFULL}\\
&&{\textstyle \frac{1}{2}}\left[g^{jk}(g_{ij,u}+e_{j,i}-e_{i,j} -f_i e_j)\right]_{,k}
 +{\textstyle \frac{1}{2}}\left[g^{jk}(g_{ij,u}+e_{j,i}-e_{i,j} -f_i e_j)\right]\,(\ln \sqrt{g}\,)_{,k}\nonumber\\
&&\qquad
 -{\textstyle \frac{1}{2}}f_i \,e^j f_j  - {\textstyle \frac{1}{2}}b_{,i} + {\textstyle \frac{1}{2}} f^je_{j,i} +{\textstyle \frac{1}{2}}g^{jk}g^{lm}g_{im,k}(e_{l,j}-e_{j,l}+f_l\,e_j-f_j\,e_l) \nonumber\\
&&\qquad {\textstyle
+\frac{1}{2}f_{i,u}+\frac{1}{2}f_i(\ln \sqrt{g}\,)_{,u} - (\ln \sqrt{g}\,)_{,ui}-\frac{1}{4}g^{jk}g^{lm}g_{km,i}\,g_{jl,u}} \label{gEinstein_equation_ui_order_zeroFULL}\\
&&\qquad =
\left(\frac{2\Lambda}{D-2} -  \frac{\etta^2-2\q^2}{D-2} \right)\!e_i + 2\q\, \xi_i - 2 F_{ij}(\q \,e^j + g^{jk} \xi_k)\,.\nonumber
\end{eqnarray}

Moreover, from the conditions  ${f_{i,j}-f_{j,i}=0}$ it follows that there exists (at least in some neighbourhood) a ``potential'' function ${{\cal F}(x,u)}$ such that $f_i$ can be written as a gradient,
\begin{equation}
f_i=\frac{\partial\cal F}{\partial x^i}\, .
\end{equation}
It is possible to employ the coordinate freedom (\ref{free3}) to achieve (say) ${f_1\not=0}$ and ${f_i=0}$ for all ${i\not=1}$. Indeed, ${\cal F}$ itself can be taken as a new local coordinate ${\tilde x^1 \equiv {\cal F}(x,u)}$, and  relation (\ref{free3}) thus implies
${\tilde f_1=1}$ and ${\tilde f_2=\tilde f_3=\dots=0}$. Therefore, without loss of generality, we can (locally) assume
\begin{equation}
f_1=1\,, \quad f_i=0 \quad \hbox{for}\quad i=2, 3, \dots, D-2 \, .
\label{canoncho}
\end{equation}
This facilitates further simplification since, for example, any derivative of $f_i$ in (\ref{gR_ij_tensor})--(\ref{gEinstein_equation_ui_order_zeroFULL}) vanishes identically.

\section{Important subfamilies and relation to previous works}
\label{previous}

In this final section we will briefly mention some particular subclasses of the metric (\ref{gen_metric gen}) and list the related references.

\subsection{pp-waves}
\label{ppwaves}
One important subclass of Kundt spacetimes are \emph{pp-waves}. These are defined geometrically as admitting a \emph{covariantly constant null vector} field $\boldk$ (and thus are sometimes denoted as CCNV spacetimes). Since
${k_{\alpha;\beta}=\frac12g_{\alpha\beta,r}}$, it follows that all the metric functions $g_{\alpha\beta}$ in (\ref{gen_metric gen}) must be independent of the coordinate $r$, so that ${f_i=0}$ for all $i$ (implying ${\tau_i=0}$), and also ${\A=0=b}$. This is thus a special case of (\ref{gen_metric fnula}),
\be
 \d s^2=g_{ij}\,\d x^i\d x^j+2\,e_i\,\d x^i\d u-2\,\d u\,\d r+ c\,\d u^2 ,\,
 \label{gen_metric pp}
\ee
where, due to (\ref{sDEF AA}),
\begin{equation}\label{sDEF AAconseq}
\etta^2+2(D-3)\q^2= 2\Lambda \ge 0  \,.
\end{equation}
Also, using the fact that $\q, F$ and ${\,}^{\rm s\!}R$ can only be functions of $u$, see (\ref{sEinstein_equation_uu_order_two}), (\ref{sEinstein_equation_ui_order_oneFULL}) and the trace of (\ref{sR_ij_tensor}), the field equations become
\begin{eqnarray}
&&{\,}^{\rm s\!}R_{ij} = \frac{2\Lambda}{D-2} \, g_{ij} - \frac{\etta^2 - 2 \q^2 }{D-2}\,g_{ij}+2 F_{ik} F_{jl}\, g^{kl}\,,\label{sR_ij_tensorpp}\\
&&{\textstyle \frac{1}{2}}(g^{ij}c_{,j})_{||i} -{\textstyle\frac{1}{2}}g^{ij}g^{kl}e_{j,l}(e_{i,k}-e_{k,i})-(g^{ij}e_{j,u})_{||i}\nonumber\\
&&\qquad{\textstyle  +\frac{1}{4}g^{ij}g^{kl}g_{ik,u}\,g_{jl,u} + (\ln \sqrt{g}\,)_{,uu}} = {\textstyle  -2(\xi_i + \q\, e_i) (\xi^i + \q\, e^i)}\,,\label{sEinstein_equation_uu_order_zeropp}\\
&&{\textstyle \frac{1}{2}}\left[g^{jk}(g_{ij,u}+e_{j,i}-e_{i,j})\right]_{,k}
 +{\textstyle \frac{1}{2}}\left[g^{jk}(g_{ij,u}+e_{j,i}-e_{i,j})\right]\,(\ln \sqrt{g}\,)_{,k}\nonumber\\
&&\qquad + {\textstyle \frac{1}{2}}g^{jk}g^{lm}g_{im,k}(e_{l,j}-e_{j,l})  {\textstyle
 - (\ln \sqrt{g}\,)_{,ui}-\frac{1}{4}g^{jk}g^{lm}g_{km,i}\,g_{jl,u}} \label{sEinstein_equation_ui_order_zeroFULLpp}\\
&&\qquad =
\left(\frac{2\Lambda}{D-2} -  \frac{\etta^2-2\q^2}{D-2} \right)\!e_i + 2\q\, \xi_i - 2 F_{ij}(\q \,e^j + g^{jk} \xi_k)\,.
\nonumber
\end{eqnarray}

In \emph{vacuum}, we have ${F_{ij} =0 =\q\,}$, so that ${\etta^2=0}$, and also ${\xi_i=0}$. In view of (\ref{sDEF AAconseq}), ${\Lambda=0}$ and due to (\ref{sR_ij_tensorpp}) the transverse Riemannian space must be Ricci flat,
\be
{\,}^{\rm s\!}R_{ij} = 0\,.
\ee
Of course, the right-hand sides of equations (\ref{sEinstein_equation_uu_order_zeropp}) and (\ref{sEinstein_equation_ui_order_zeroFULLpp}) are also zero.

In the particular case when all the functions ${e_i}$ can be \emph{globally} removed, i.e., ${f_i=0=e_i}$, equations (\ref{sEinstein_equation_uu_order_zeropp}), (\ref{sEinstein_equation_ui_order_zeroFULLpp}) further reduce to a much simpler form
\begin{eqnarray}
&&{\textstyle  \frac{1}{2}(g^{ij}c_{,j})_{||i} + \frac{1}{4}g^{ij}g^{kl}g_{ik,u}\,g_{jl,u}+ (\ln \sqrt{g}\,)_{,uu}} =
{\textstyle  -2\,\xi_i\,\xi^i}\,,\label{sEinstein_equation_uu_order_zeropps}\\
&&{\textstyle \frac{1}{2}}\left[g^{jk}(g_{ij,u})\right]_{,k} +{\textstyle \frac{1}{2}}\left[g^{jk}(g_{ij,u})\right]\,(\ln \sqrt{g}\,)_{,k}
 {\textstyle - (\ln \sqrt{g}\,)_{,ui}-\frac{1}{4}g^{jk}g^{lm}g_{km,i}\,g_{jl,u}} \label{sEinstein_equation_ui_order_zeroFULLpps}\\
&&\hskip67mm =  2\q\, \xi_i - 2 F_{ij}g^{jk} \xi_k\,.
\nonumber
\end{eqnarray}

The pp-wave spacetimes (\ref{gen_metric pp}) in higher dimensions were introduced in the classic paper \cite{Bri25} by Brinkmann, and since then they have been studied extensively, see for example \cite{Ort04,ColMilPraPra04,ColMilPelPraPraZal03,ColFusHerPel06,Col08} and references therein.

\subsection{VSI and CSI spacetimes}
\label{VSI}

Higher-dimensional Lorentzian spacetimes with \emph{vanishing} scalar curvature invariants of all orders (the so-called VSI spacetimes) were explicitly presented in \cite{ColFusHerPel06}, see also \cite{ColMilPelPraPraZal03,Coleyetal04,ColMilPraPra04,ColHerPel06,Col08}. It was found that all such spacetimes belong to the Kundt class and, in fact, can be written in the canonical form
\be
 \d s^2=\delta_{ij}\,\d x^i\d x^j+2\,(e_i+ f_i\,r )\,\d x^i\d u-2\,\d u\,\d r+(\A\,r^2 + b\,r + c)\,\d u^2 \,.
 \label{metricVSI}
\ee
This is a particular case of the metric (\ref{gen_metric gen}) in which the
\emph{transverse space is flat\,}, i.e.
\be
 g_{ij}=\delta_{ij}\,.
 \label{flatspace}
\ee
Such spacetimes have a Weyl tensor of algebraic type~III, or more special. Two subclasses can be distinguished, namely the case ${f_i=0}$ for all ${i=1,2,\ldots, D-2}$ (which occurs when ${\,f^i\! f_i=0\,}$) and the case ${f_1\not=0}$ with ${f_i=0}$ for ${i=2, 3, \dots, D-2}$  (when ${\,f^i\! f_i>0\,}$). For more details, see \cite{ColFusHerPel06} with the notational identification ${r=-v}$, ${e_i=W_i^{(0)}}$ and ${f_i=-W_i^{(1)}}$ where either ${f_1=0}$ or ${f_1=2/x^1}$. Note that the special choice (\ref{canoncho}) in which ${f_1=1}$ can be achieved by a simple coordinate transformation ${x^1\to\exp(\frac12 x^1)}$ with redefinition ${e_i=\frac12 W_i^{(0)}\exp(\frac12 x^1)}$, of course at the expense that the flat metric $\delta_{ij}$ becomes ${\,\hbox{diag}\,(\frac14\exp x^1, 1,\ldots,1)}$.

An important subclass of VSI spacetimes with ${f_i=0}$ for all~${i}$ are pp-waves with a flat spatial metric (\ref{flatspace}), as described above in section~\ref{ppwaves} (see also the following section~\ref{gyratons}).

A generalisation of the VSI spacetimes belonging to the Kundt class, such that all polynomial scalar invariants constructed from the Riemann tensor and its derivatives are \emph{constant} (denoted CSI spacetimes), was introduced and studied in \cite{ColHerPel06,Col08}.

\subsection{Gyratons}
\label{gyratons}

Another physically interesting subclass of the Kundt family of non-expanding space-times are the so-called \emph{gyratons}. These describe the field of a localised spinning source that propagates at the speed of light. Such a situation may be modelled by a spinning beam pulse of radiation (or null matter) that has a finite duration in retarded time~$u$ and a negligible transverse radius, so that the total energy and angular momentum remain finite. Exact spacetimes of this type, represented by specific axially symmetric type~III Kundt solutions in ${D=4}$, were introduced by Bonnor \cite{Bon70} (who called them ``spinning nullicons'') and recently generalised to higher dimensions \cite{FroFur05,FroIsrZel05,FroZel05,FroZel06,FroLin06,CalLeZor07}, see also \cite{ColFusHerPel06,Col08}.

The \emph{external} gravitational field of the simplest gyraton with ${\Lambda=0}$ is described by the metric
\be
 \d s^2=\delta_{ij}\,\d x^i\d x^j+2\,e_i\,\d x^i\d u-2\,\d u\,\d r+ c\,\d u^2 ,\,
 \label{simple gyraton}
\ee
see \cite{FroIsrZel05} with the identification ${e_i=A_i}$, ${c=\Phi}$. Obviously, this is a particular pp-wave Brinkmann metric (\ref{gen_metric pp}) with ${g_{ij}=\delta_{ij}}$, ${e_i\not=0}$, and also a special subfamily of VSI spacetimes (\ref{metricVSI}) with ${f_i=0}$ and ${a=0=b}$. The metric functions $e_i$ and $c$ must satisfy \emph{vacuum} field equations (\ref{sEinstein_equation_uu_order_zeropp}) and (\ref{sEinstein_equation_ui_order_zeroFULLpp}) with vanishing right-hand sides. Interestingly, these are formally equivalent to equations for the magnetic vector potential ${A_i}$ and electric scalar potential ${\Phi}$ in the transverse ${(D-2)}$-dimensional flat space \cite{FroFur05,FroIsrZel05}, and thus can be generally solved by standard methods.

In order to obtain the complete spacetime, it is necessary to find also the corresponding \emph{internal} solution inside the gyraton, and match it to the vacuum exterior solution (\ref{simple gyraton}). It must be emphasised that the ``gyraton matter'' is  spinning, so that its energy-momentum tensor has not only the familiar pure radiation (null fluid) component ${T_{uu}\not=0}$  but also an extra non-diagonal term ${T_{ui}\not=0}$ (other components, and thus the trace ${T_\mu^{\, \mu}}$,  are assumed to be zero). In the natural frame (\ref{frame}), the only non-vanishing components of the Ricci tensor, related to the gyraton, are thus ${\Phi\equiv R_{\alpha\beta}\,l^\alpha\, l^\beta}$ and ${\Phi_i\equiv R_{\alpha\beta}\,l^\alpha\, m^\beta_{(i)}}$, which in ${D=4}$ correspond to the Newman--Penrose scalars ${\Phi_{22}}$ and ${\Phi_{12}}$, respectively. Within this internal gyraton region, the  spacetime is in general of algebraic type III.

As discussed in section~\ref{coordfreedom}, it is possible to remove all metric functions $e_i$ in the exterior vacuum region, but only locally. In the presence of gyratonic matter, we \emph{can not set} ${e_i=0}$ \emph{globally} because the exterior pp-wave manifold (\ref{simple gyraton}) is not simply connected and even if the spinning gyraton source is negligibly small, there remains a particular singularity along (part of) the  axis ${x^i=0}$.

Further specific generalisations of the gyraton spacetime (\ref{simple gyraton}) have been presented recently which admit a negative cosmological constant \cite{FroZel05} or a charged source \cite{FroZel06}. Gyraton solutions in supergravity theories have been considered in \cite{FroLin06,CalLeZor07}.

All these gyratonic solutions belong to the Kundt family. Naturally, other types of gyratons can be explicitly constructed and identified within the general family of higher-dimensional Kundt spacetimes presented in this contribution. For example, in \cite{ColFusHerPel06} the authors mention the existence of both pp-wave gyratons (with no $r$-dependence) and the Kundt gyratons (with $r$-dependence) which generalise a metric of the Kundt waves. Obviously, the latter can be identified in the canonical subclass of spacetimes with ${\,f^i\! f_i>0\,}$, described in section~\ref{subclasse2}.

\section{Conclusions}
\label{conclusions}

We presented and discussed the main features of the general Kundt family of higher-dimen\-sional spacetimes, which admit a hypersurface-orthogonal, non-shearing and non-expanding congruence of null geodesics. In particular, without assuming any specific matter content, we explicitly calculated all components of the curvature and Ricci tensors for the Kundt metric. We also determined its algebraic type, together with constraints imposed by general Einstein's field equations.

Starting from section~\ref{EinsteinMaxwell}, we restricted our analysis to the most important case of vacuum Kundt spacetimes with an arbitrary cosmological constant $\Lambda$, and possibly with an aligned electromagnetic field. We derived the explicit form of all Einstein(--Maxwell) equations within such a setting. The results are summarised in section~\ref{summary} where we also discuss the remaining coordinate freedom and define canonical subclasses.

We demonstrated above that the \emph{general} form of such (electro)vacuum Kundt spacetimes in any dimension $D$ is simple, see the metric (\ref{gen_metric gen}). In particular,
the metric functions $g_{ui}$ are at most linear while $g_{uu}$ is at most quadratic in the affine parameter~$r$ of the congruence generated by the multiple WAND ${\partial_r}$, which fully agrees with a form of vacuum non-twisting metrics presented in \cite{PraPra08}. However, the structure of the remaining Einstein(--Maxwell) equations, which determine the \emph{explicit} form of the metric functions, is rather complicated. Obviously, these equations are not manageable in general. This is not, in fact, surprising since even in the ${D=4}$ case such solutions are mostly not known explicitly, see \cite{Stephanibook}. Nevertheless, various canonical and special subclasses of vacuum and electrovacuum Kundt spacetimes, possibly with a non-vanishing cosmological constant~$\Lambda$, can be identified and studied. These include some previously found solutions in this family, namely generalised pp-waves, VSI or CSI spacetimes, and gyratons, as described in section~\ref{previous}.

In contrast to the Robinson--Trautman family of expanding solutions, which in ${D>4}$ contains only conformally flat and type~D solutions~\cite{PodOrt06,OrtPodZof08}, the non-expanding class of higher-dimensional Kundt spacetimes is very rich. There are vacuum and electrovacuum solutions of various algebraic types (i.e., of type~II and more special), and the transverse ${(D-2)}$-dimensional Riemannian space admits many spatial metrics. This follows from the property that Kundt's spacetimes have a shear-free and non-expanding character and thus, in ${D>4}$, they are closer to their four-dimensional counterparts, see~\cite{ColHerPel06} and references therein.

\section*{Acknowledgments}

The authors are grateful to Marcello Ortaggio, Vojt\v ech Pravda and Alena Pravdov\'a for their comments on the manuscript. This work was supported by the grant GA\v{C}R~202/08/0187 and by the Czech Ministry of Education under the projects MSM0021610860 and LC06014.




\end{document}